# Orbitronics: Light-induced Orbit Currents in Terahertz Emission Experiments

Yong Xu[a,b], Fan Zhang[a,b], Albert Fert[a,c], Henri-Yves Jaffres[c], Yongshan Liu[a,b], Renyou Xu[a,b], Yuhao Jiang[a], Houyi Cheng[a,b], and Weisheng Zhao[a,b]

[a]MIIT Key Laboratory of Spintronics, School of Integrated Circuit Science and Engineering, Beihang University, Beijing 100191, China
[b]Hefei Innovation Research Institute, Beihang University, Hefei 230013, China
[c]Unité Mixte de Physique, CNRS, Thales, Université Paris-Saclay, Palaiseau 91767, France
[1]These authors contributed equally: Yong Xu, Fan Zhang, Albert Fert, Henri-Yves Jaffres
[2]Corresponding author: albert.fert@cnrs.fr, weisheng.zhao@buaa.edu.cn

**Orbitronics is based on the use of orbit currents as information carriers. Up to now, orbit currents were created from the conversion of charge or spin currents, and inversely, they could be converted back to charge or spin currents. Here we demonstrate that orbit currents can also be generated by femtosecond light pulses on Ni. In multilayers associating Ni with oxides and nonmagnetic metals such as Cu, we detect the orbit currents by their conversion into charge currents and the resulting terahertz emission. We show that the orbit currents extraordinarily predominate the light-induced spin currents in Ni-based systems, whereas only spin currents can be detected with CoFeB-based systems. In addition, the analysis of the time delays of the terahertz pulses leads to relevant information on the velocity and propagation of orbit carriers. Our finding of light-induced orbit currents and our observation of their conversion into charge currents opens new avenues in orbitronics, including the development of orbitronic terahertz devices.**

The conversion from a charge current $j_C$ to a spin current $j_S$ has been studied extensively in systems with strong spin-orbit coupling [1,2]. The most successful example of the conversion mechanism is the spin Hall effect (SHE) for nonmagnetic metals containing heavy atoms [3–6]. For two-dimensional electron gases at Rashba interfaces and the surfaces or interfaces of topological insulators, the charge-to-spin conversion is usually named the Rashba-Edelstein effect (SREE), with S for spin in our notation) [7–10]. In ferromagnet (FM)/nonmagnet (NM) heterostructures, the spin currents induced by SHE or SREE are strong enough to reverse the magnetization of the FM material by Spin Transfer Torque [11,12]. Therefore, both SHE and SREE have attracted much attention due to their technological significance in developing future magnetic memory devices. Their inverse effects, namely, inverse SHE (ISHE) [13] and inverse SREE (ISREE) [14], convert a spin current $j_S$ into a charge current $j_C$. These effects have been widely utilized for detecting spin currents generated by other stimuli, such as the heat current [15], the charge current [5,6], and the spin-pumping technique [16–18]. In recent years, ISHE and ISREE have also been utilized to generate ultrafast charge pulses and develop efficient broadband spintronic terahertz emitters [19–22].

Several recent works have highlighted the importance of the orbital degree of freedom in condensed matter physics and kicked off the emergent research field of orbitronics [23]. Orbitronics exploits the transport of orbital angular momentum through materials by orbit currents which can be exploited as an information carrier in solid-state devices. As for the conversion between spin and charge current by SHE or SREE, it has been theoretically predicted and experimentally shown that a charge current $j_C$ can be converted into an orbit current $j_L$ via the orbital Hall effect (OHE) or the orbital Rashba-Edelstein effect (OREE) [24–31].

Unlike spin, the orbit current cannot exert a torque directly on magnetization due to the lack of direct coupling between the orbit current and the magnetization. However, the spin-orbit coupling can convert the orbit current $j_L$ into spin current $j_S$, thereby generating torques on the magnetization. Several studies have identified the spin torques originating from OHE or OREE in heterostructures [29,30].

According to the Onsager reciprocal relation, the inverse effect of the OHE or OREE should convert an orbit current into a charge current. To date, few studies reported the conversion from orbit current to charge current. This is due to the lack of a reliable orbit current source and the disturbance of the omnipresent ISHE. In this paper, we provide strong evidence for conversion from a current orbital $j_L$ into a charge current through free-space terahertz emission. The new effects, the orbital counterpart of ISHE and ISREE, is named the inverse orbital Hall effect (IOHE) and the inverse orbital Rashba-Edelstein Effect (IOREE) [23]. Thanks to the induced charge current, the IOHE and IOREE provide a feasible approach for the electrical characterization of the orbit current.

Here, we present experiments of THz emission in the time domain generated by the conversion of light-induced spin and orbit currents into ultrafast charges currents in MgO/NM/F multilayers in which NM = Pt, Ta or Cu and F = CoFeB or Ni (fabrication and experimental details in Methods). We show that, with Ni, the main contribution to the THz emission originates from orbit currents and can be analyzed to derive information on the velocity and propagation of orbit carriers.

**Identification of orbital-to-charge conversion**

We begin by comparing THz waveforms from MgO(3 nm)/NM (4 nm)/Co$_{0.2}$Fe$_{0.6}$B$_{0.2}$ (10 nm)/MgO/Ta and MgO(3 nm)/NM (4 nm)/Ni (10 nm)/MgO/Ta, where NM = Ta, Pt, or Cu (Figure 1). The NM/CoFeB samples follow the typical behavior of spintronic terahertz emission driven by conversion from spin currents into charge currents via ISHE in NM. The ultrafast charge current induced by the spin current is given as $j_C = \gamma_S j_S$, where $j_S$ refers to the spin current, and $\gamma_S$ is the spin-to-charge efficiency of the NM layer. For a given sample, the waveform polarity shows the expected reversal when the sample is flipped from front excitation to back excitation. Moreover, the waveform polarities of NM/CoFeB, dictated by the sign of the SHE of NM, are opposite for Ta relative to Pt, as naturally expected from their opposite SHE [4, 5]. In agreement with the very small spin-orbit interaction and negligible ISHE in Cu, the sample CoFeB/Cu shows a small signal very similar to the signal obtained with a CoFeB monolayer without any evidence of additional contribution from Cu. The intrinsic signals in CoFeB monolayers can be ascribed to the anomalous Hall effect (AHE) of the CoFeB layer [32,33], as it will be discussed later.

In contrast, the waveforms of NM(4 nm)/Ni(10 nm) show reversed polarities with respect to Ni monolayer and the same polarities for all three samples, regardless of the sign of spin Hall angle of the NM layers (Figure 1b). Due to the opposite spin Hall angle of Pt and Ta, spin-charge conversion via ISHE should give opposite polarities for Pt/Ni and Ta/Ni. In addition, as Cu(4 nm)/Ni(10 nm) shows a waveform opposite to the intrinsic one of the Ni monolayer, this reversal reveals a significant additional contribution from Cu in spite of its minimal spin Hall angle. Our results with Ni, the same polarity with Pt and Ta and significant contribution from the addition of Cu, show that ISHE is not sufficient to explain the THz waveforms for Ni-based bilayers.

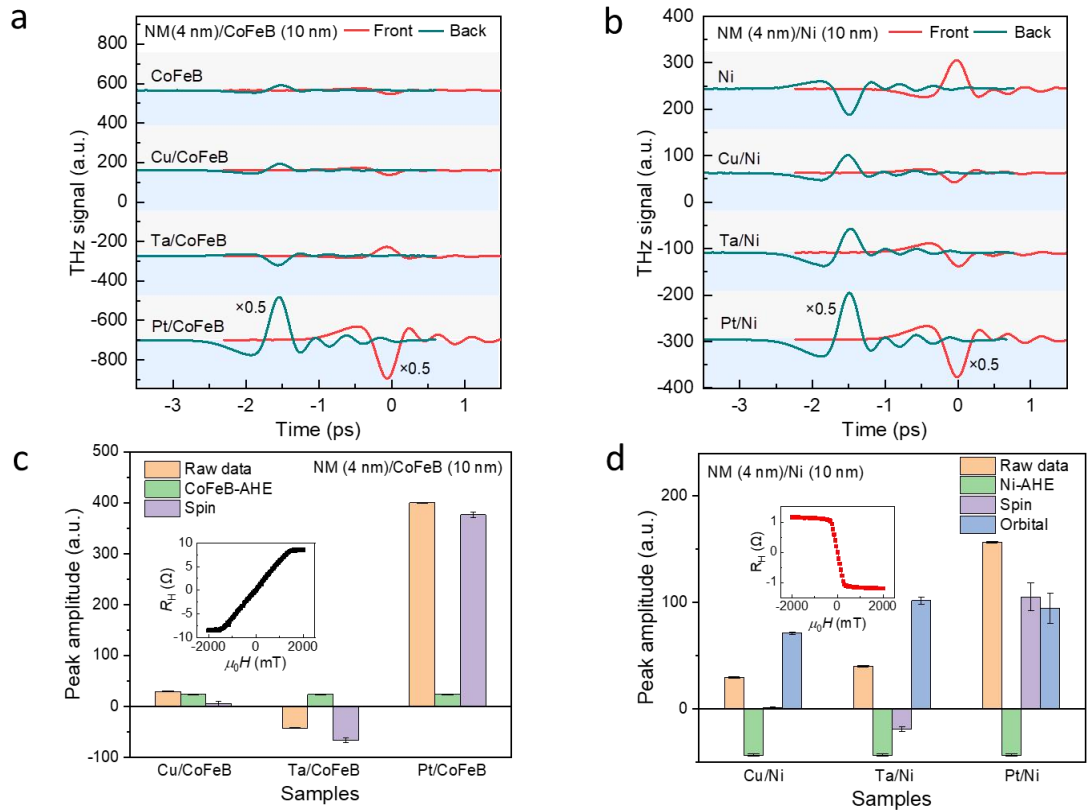

Figure 1 A comparison between THz signals from NM/CoFeB and NM/Ni

heterostructures. a) THz waveforms emitted from CoFeB (10 nm), Cu (4 nm)/CoFeB (10 nm), Ta (4 nm)/CoFeB (10 nm), and Pt (4 nm)/CoFeB (10 nm) measured with front and back sample excitations. b) THz waveforms emitted from Ni (10 nm), Cu (4 nm)/Ni (10 nm), Ta (4 nm)/Ni (10 nm), and Pt (4 nm)/ Ni (10 nm) measured with front and back sample excitation. An in-plane magnetic field of 80 mT is applied. The above experiments were carried out under the same experimental conditions. c-d) Schematic illustration of the decomposition of the back THz signals of NM/CoFeB in panel a) or NM/Ni in b) into the sum of the signals in CoFeB or Ni single layers (we call it AHE, see later) and the added contribution introduced by NM in bilayers (we call it spin + orbital). The insets display the Hall resistance $R_H$ as a function of the out-of-plane magnetic fields $H$ for CoFeB and Ni layers in insets of c) and d), respectively. Their opposite AHE explains the opposite emissions by CoFeB and Ni single layers.

Since the results of NM/Ni cannot be fully explained with the spin-to-charge conversion by ISHE, we examine the other possible mechanisms. Excluding magnetic dipole emission (first reported by Beaurepaire et al. [34]) which cannot give opposite polarities for front and back excitations, we consider THz emission driven by AHE, and THz emission driven by spin-to-charge conversion or orbit-to-charge conversion.

Considering the AHE contribution, the intrinsic emissions by Ni and CoFeB monolayers are predominantly related to AHE with opposite polarities in agreement with the opposite AHE of Ni and CoFeB shown in the inset of Fig.1c-d. In addition, it is important to note in Fig.1a that only adding metals with significant SHE (Ta/CoFeB, Pt/CoFeB) changes the signal of the CoFeB monolayer, while adding 4 nm of Cu changes it in negligible way, as we also checked with other thicknesses of Cu in our thickness range. It means that the contribution from AHE in a ferromagnetic layer is negligibly affected by the addition of a nonmagnetic metallic layer as Cu and, for example, can be supposed to be independent of the Cu thickness in the Cu/Ni discussed later.

The spin-to-charge conversion explains THz emission in NM/CoFeB bilayers. However, the THz emissions of the NM/Ni bilayers are in contradiction with this mechanism in the following aspects: (i)The addition of Ta, Pt, or Cu leads to significant emissions of the same polarity for all three nonmagnetic metals, and this polarity is reversed with respect to the polarity of the AHE-induced emission by Ni monolayers without NM. (ii)Since the spin Hall angle is very small for Cu, this reversal with respect to the AHE-induced signal in the Ni monolayer cannot be explained by ISHE. (iii)The same polarity of the signals for Ta and Pt is also in contradiction with the conversion of spin current to charge current since Ta and Pt have opposite SHEs.

As in very recent THz results with Ni [35], our results can be explained by a significant light-induced generation of orbit currents in Ni, as illustrated by Figure 2a. Among the contributions to THz emission, the charge currents generated by orbit-to-charge conversion of these orbit currents in the NM layer dominates over both the AHE intrinsic signal of Ni and the spin-to-charge conversion of the light-induced spin currents. Quantitatively we can write for the resulting charge current:

$$j_C = \gamma_{AHE} j_l + \gamma_L j_L + \gamma_S j_S$$

where $\gamma_L$ and $\gamma_S$ are orbit-to-charge and spin-to-charge conversion coefficients. The conversion of the orbit and spin current emitted from the ferromagnetic layer can occur inside the nonmagnetic layer (IOHE or ISHE) or at the interface of the nonmagnetic layer with, in our sample, MgO (as illustrated by Fig.2a). Our results for NM/Ni, with the polarities for Cu, Ta, and Pt aligned in the same direction (opposite to the direction for a Ni single layer) show the predominance of the orbital terms (bulk IOHE and interfacial IOREE) on spin terms (ISHE and ISREE) and AHE terms, finally aligning all the polarities. As we have seen, the results with CoFeB present the different signs expected for ISHE in Pt and Ta, and also the absence of a significant signal added by Cu in Cu/CoFeB, in agreement with the very small ISHE expected for Cu. They indicate that CoFeB is much less efficient than Ni for the production of light-induced orbit currents.

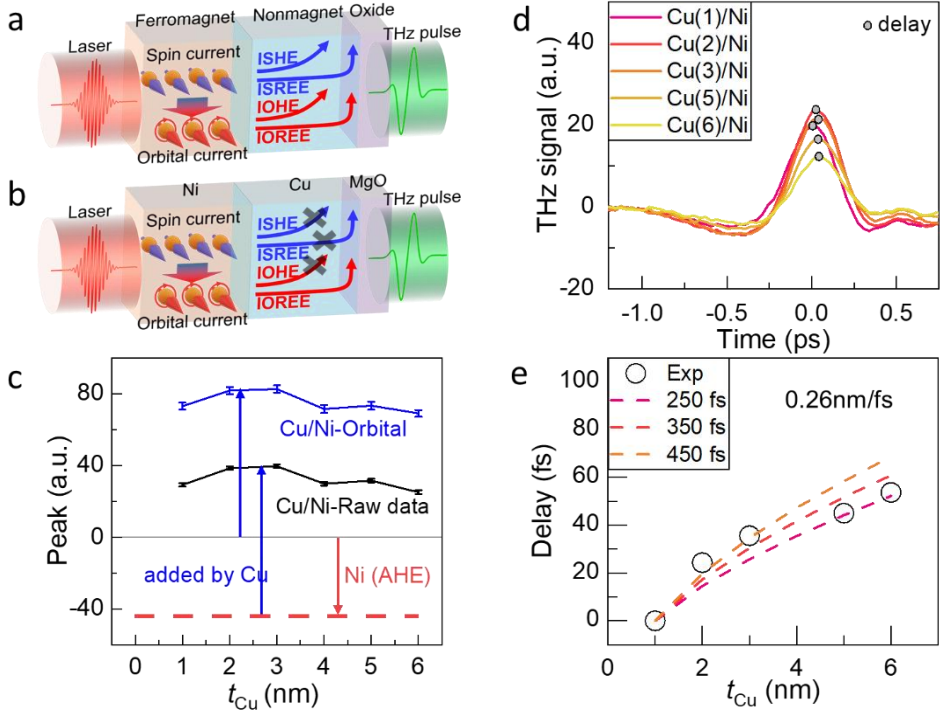

Figure 2: a) Conceptional diagram of the THz emission showing the different channels for conversion from spin and orbit currents emitted by the ferromagnet into charge currents in the nonmagnetic layer and at the interface with MgO. b) Same diagrams as in a) for the situation of Cu/Ni bilayer. The crosses indicate the conversion channels which are not involved in the THz emission. c) THz peak amplitudes for substrate/MgO/Cu/Ni(10 nm) samples with raw data and orbital contribution as a function of the Cu thickness. The amplitude of the orbital contribution is obtained by removing from the raw amplitude the intrinsic AHE contribution of Ni single layer supposed to be independent of Cu thickness (at least in first approximation), as justified in the text. d) THz emission signal for substrate//MgO/Cu/Ni (10 nm) samples for different Cu thicknesses. The maxima of the signals are highlighted by a circular marker. e) Delay $\tau_D$ versus Cu thickness from experimental results in Fig.2 d. Circles indicate error bars. Curves are calculated from Eq.2 with the parameters $v_F = 0.26$ nm/fs and three values of $\tau_{of}$, see inset.

**Analysis of THz emission induced by orbital currents in Cu/Ni bilayers.**

The analysis of the orbital contributions is simpler for MgO/Cu/Ni for several reasons: i) no or negligible ISHE term due to the very small ISHE in Cu, ii) no ISREE contribution at MgO/Cu interface because it should have appeared in MgO/Cu/CoFeB in addition to the signal with only MgO/CoFeB, iii) AHE contribution negligibly dependent on Cu thickness as supported by the negligible departure from the signal with CoFeB monolayer in Cu/CoFeB samples with different Cu thicknesses in our thickness range (Supporting information and referred to above). The only contributions to consider are a quasi-thickness-independent AHE contribution and those induced by orbit current, IOHE in Cu, and IOREE at Cu/MgO, as illustrated by Fig.2b with crosses on the ISHE and ISREE channels (the additional cross on the IOHE channel will be discussed below)

In Fig.2c, we show the dependence of the peak amplitude for Cu/Ni samples as a function of Cu thickness, with, in blue, the signal ascribed to orbit current by removing the AHE contribution (assumed to be constant) from the raw data. The signal for Cu/Ni starts as soon as the thinnest Cu layer (1nm) is deposited to cover almost totally MgO and then remains almost at about the same level within the thickness range. A conversion of the orbit current by IOHE in the bulk of the Cu layer would lead to a progressive increase of the signal following the increase of active region for IOHE at the increasing thickness of Cu. In addition, we do not see some increase in the width of the waveform as would be expected with contributions coming from different depths in a Cu layer and differently delayed in time. The thickness dependence in Fig.2c is characteristic of THz emission by orbit to charge conversion by IOREE at the MgO/Cu interface with the expected additional slow decrease with thickness due to increasing light absorption

Recent experiments of THz emission by Ni/W bilayers have also revealed the light-induced emission of orbit currents by Ni [35]. Their interfacial conversion by IOREE at W/CuOx interfaces introduces time delays $\tau_D$ time shifts) in the waveforms of THz emission, and interesting information can be derived from the variation of these delays at increasing W thickness, from almost linear in the ballistic regime to quadratic in the diffusive regime.

The observation of an almost linear variation of the delay $\tau_D$ in the inset of Fig.2d, as well as a clear concave shape variation at the smaller thickness $t_{Cu}$, is thus consistent with the ballistic regime of orbital injection. The typical average slope is given by the orbital group velocity of carriers $v_F \approx 0.26$ nm/fs, which is an order of magnitude below the charge carriers velocity in Cu at room temperature. This striking result is not really surprising because charge currents and orbit currents cannot be carried by the same type of carriers. In the particular case of Cu, one knows that the charge current is predominantly carried by s-band electrons (no orbital moment) whereas orbit currents are necessarily carried by other bands (and at higher energy for light-induced carriers).

Another striking result observed in the inset of Fig.2d is then the clear concave shape of $\tau_D$ vs. $t_{Cu}$ as an opposite trend of the $t_{Cu}$-dependence of $\tau_D$ expected from diffusion processes. Such behavior is observed when an orbital-flip probability $p_{of} = \frac{\tau_D}{\tau_{of}}$ ($\tau_{of}$ is the corresponding orbital flip time possibly due to orbital decoherence) is considered during the ballistic transport from one interface to the other (see the supplementary information for the calculation details), and this can be understood as follow. For a ballistic transport the typical cone aperture for the transport in k-space is given by $\cos\theta_k = \frac{t_{Cu}}{V_F \tau_{of}}$ , as the result that the average transport cone angle $\theta_k$ increases when $t_{Cu}$ decreases leading, in parallel, to an apparent reduction of the apparent group velocity through the increase of the travelling distance (concavity of the curve). Conversely, a very short orbital flip time $\tau_{of}$ would lead to quasi-normal injection ($\theta_k \approx 0$) and a quasi $\tau_D$ vs. d linear dependence ($\tau_D = \tau_0 = \frac{d}{V_F}$). The observation of a certain concavity shape is then the demonstration of a pretty large value of $\tau_{of}$ with a well smaller role played by the scattering momentum time $\tau$. The result of the calculation (see supplementary information):

$$\tau_D = \frac{\int_{t_o}^{\infty} d\tau\, exp\left(-\frac{t}{\tau_{of}}\right)}{\int_{t_o}^{\infty} \frac{d\tau}{\tau} exp\left(-\frac{t}{\tau_{of}}\right)}$$

and the fit of the experimental delays in Fig.2e leads to a reasonable agreement with $v_F = 0.26$ nm/fs and $\tau_{of} = 350$ fs.

To sum up, the analysis of the results in the simplest situation of MgO/Cu/Ni leads to two main results, 1) THz emission due predominantly to conversion of orbit current by IOREE at MgO/Cu interface. 2) Interesting data on the dynamics of the orbit carriers emitted by Ni. In Section 2 of Supplementary Information, we show how the results on $v_F$ and $\tau_{of}$ can be used to derive the orbit diffusion length of the propagation of the orbital polarization from carefully driven THz emission experiments.

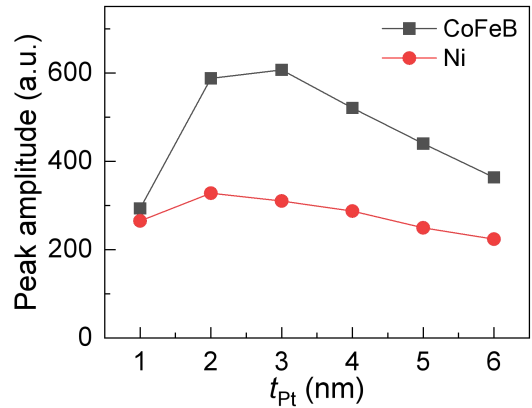

Figure 3: THz peak amplitudes from raw data of Ni (5 nm)/Pt and CoFeB (5 nm)/Pt as a function of the Pt thickness measured with front excitation.

**THz emission combining contributions from spin and orbit currents.**

We now consider THz emission in systems with added contributions from conversions of both spin currents (by ISHE and ISREE) and orbit currents (by IOHE and IOREE), as in the example of samples with Pt for the NM layer and CoFeB or Ni for the F layer in Fig.3. With CoFeB, as shown by the opposite signals in Fig.1a for opposite signs of ISHE in Pt and Ta or W, one knows that a predominant contribution comes from ISHE. The THz signal with CoFeB in Fig.3 increases rapidly with Pt thickness up to a maximum at 3nm of Pt, which is consistent with the progressive increase of the ISHE active region at increasing thickness. The slow decrease after the maximum is what is expected from the increase of light and THz absorption. With Ni, the dependence on Pt thickness (red curve) is different and, qualitatively, appears as a variation with thickness similar to what is observed with CoFeB (black curve) on top of a plateau reminding the plateau observed with Cu/Ni in Fig.2c and ascribed to the interfacial conversion of the orbit current emitted by Ni. Precise separation of the contributions of the different conversion mechanisms, ISHE, ISHEE, IOHE, and IOREE illustrated in Fig2b would be possible with large series of new types of samples in which the ratio between spin and orbit current can be quantitatively controlled for example, as in [36,37] by insertion of the rare-earth ultra-thin layer to induce conversions between spin and orbit current and varying their respective values, what is out of the scope of the present paper.

**Discussion**

The usual mechanisms of the generation of orbit currents are conversions from charge currents by OHE or OREE or conversions from spin currents. The most striking result of our work is the light-induced generation of orbit current in Ni. We could not find any theoretical work predicting production of orbit current by Ni from high-energy excitation. For low energy electrons, according to the coefficients calculated for the conversion between orbit and spin current in Fig.2b in [29], we can see that the orbit-to-spin conversion is definitely more efficient in Ni compared to other 3d metals or alloys, what shows some similarity with our results. For high-energy excitation, a first tentative scenario is the production of orbit current from the conversion of light-induced spin current, as described for inter-conversions between orbit and spin in the Fermi energy range [38]. In another scenario, the light-induced production of orbit current might be directly from excitation of orbital states [39]. Further theoretical works are needed to study the respective contributions of these two types of mechanism.

From an experimental point of view, the specific character of Ni for the production of orbit current is confirmed by other types of experiments. In Spin Torque FerroMagnetic Resonance (ST-FMR), Ni is described as "abnormal" from the results presented by Lee *et al* [29] in their Fig.3 on the damping-like torque on ferromagnetic metals (CoFeB, FeB, Ni) deposited on Pt or Ta (opposite SHE). Opposite torques for bilayers of CoFeB (or FeB) with Pt and Ta are observed, as expected from the opposite SHE of Pt and Ta, whereas the torque on Ni has the same sign with Pt and Ta. This striking similarity between ST-FMR and light-induced THz emission supports the idea that orbit currents can be generated not only by conversion from charge or spin currents but also by different types of excitations, including excitation by light and ferromagnetic resonance (FMR). Recently, the existence of orbit current generated by FMR of YIG ($Y_3Fe_5O_{12}$) and converted into voltage by IOREE at Pt/Cu/$O_x$ interfaces was also observed [40].

In summary, orbitronics is a promising field of research in which we have presented results obtained by a new experimental method based on the production of light-induced orbit current and their exploitation for terahertz emission in multilayers associating Ni with nonmagnetic metals NM and MgO. We find terahertz emission of the same polarity with Cu, Ta, and Pt for NM despite the opposite SHE of Ta with respect to Pt (this common polarity is also opposite to the polarity of the AHE-induced emission by a Ni monolayer). We ascribe these results to an efficient light-induced emission of orbit currents in Ni, and we have presented a detailed analysis of them in the simpler situation of MgO/Cu/Ni from which, in addition, we can extract information on the velocity and orbital flip time of the orbit carriers. The main general results are that orbit currents can be generated not only by conversion from charge or spin currents but also induced by light and, as we noted, possibly also by microwaves in experiments of FMR type. In THz emission experiments, the analysis of the emission delays can bring precious information on the dynamics of the orbit carriers. All these results open new routes for orbitronics and future orbitronic devices.

## METHODS:
**Experiment procedures**

Magnetic multilayer samples were prepared by high-vacuum magnetron sputtering on a glass substrate. Unless otherwise specified, all the samples are deposited on a 3 nm MgO buffer and protected with a capping layer of a MgO(2 nm)/Ta (2 nm) bilayer. In the THz time-domain spectroscopy, femtosecond pulses of 35-fs pulse duration, 1 kHz repetition rate, and ~1 mJ/cm$^2$ fluence are used to pump the magnetic multilayers.

**Data analyses:**

One of the strong interests of THz-TDS in the pulsed regime over other techniques is indeed its ability to probe the time dynamics of the spin and/or orbital injection process close to interfaces, presently Ni/Cu and Cu/MgO after an ultra-short pulse generation [38]. Such delay time depends on the Cu thickness ($t_{Cu}$), group velocity of carriers ($v_F$) according to $\tau_D = \frac{t_{Cu}}{V_F cos\theta}$ in the case of a ballistic transport ($t_{Cu}$ smaller than the mean free path, MFP $\lambda$). On the other hand, in the case of a diffusive transport regime, the expression is more complex $\tau_D = \frac{3t_{Cu}^2}{\lambda^2}\tau$ with $\tau$ the momentum relaxation time, leading thus to a square dependence of $\tau_D$ on the layer thickness $t_{Cu}$ [35]. Systematic measurements of the resistivity of sputtered Cu layers in a different type of multilayers and the thickness range of our samples led us to estimate the resistivity MFP to about 7 nm (Table 1 in Sup [41]), however that it should be slightly longer if we have not to consider the scattering on interfaces involved in the resistivity MFP. It means that our experiments are in the ballistic regime (at the limit of the transition to diffusive), explaining that we do not observe any square dependence of $\tau_D$ on $t_{Cu}$ in Fig.2e.

# References

[1] J. Sinova, S. O. Valenzuela, J. Wunderlich, C. H. Back, and T. Jungwirth, *Spin Hall Effects*, Rev. Mod. Phys. **87**, 1213 (2015).

[2] A. Soumyanarayanan, N. Reyren, A. Fert, and C. Panagopoulos, *Emergent Phenomena Induced by Spin–Orbit Coupling at Surfaces and Interfaces*, Nature **539**, 7630 (2016).

[3] K. Ando, S. Takahashi, K. Harii, K. Sasage, J. Ieda, S. Maekawa, and E. Saitoh, *Electric Manipulation of Spin Relaxation Using the Spin Hall Effect*, Phys. Rev. Lett. **101**, 036601 (2008).

[4] A. Hoffmann, *Spin Hall Effects in Metals*, IEEE Trans. Magn. **49**, 5172 (2013).

[5] L. Liu, C. F. Pai, Y. Li, H. W. Tseng, D. C. Ralph, and R. A. Buhrman, *Spin-Torque Switching with the Giant Spin Hall Effect of Tantalum*, Science **336**, 555 (2012).

[6] I. M. Miron, K. Garello, G. Gaudin, P. J. Zermatten, M. V. Costache, S. Auffret, S. Bandiera, B. Rodmacq, A. Schuhl, and P. Gambardella, *Perpendicular Switching of a Single Ferromagnetic Layer Induced by In-Plane Current Injection*, Nature **476**, 189 (2011).

[7] V. M. Edelstein, *Spin Polarization of Conduction Electrons Induced by Electric Current in Two-Dimensional Asymmetric Electron Systems*, Solid State Commun. **73**, 233 (1990).

[8] Y. Fan et al., *Magnetization Switching through Giant Spin–Orbit Torque in a Magnetically Doped Topological Insulator Heterostructure*, Nat. Mater. **13**, 7 (2014).

[9] A. Manchon, H. C. Koo, J. Nitta, S. M. Frolov, and R. A. Duine, *New Perspectives for Rashba Spin–Orbit Coupling*, Nat. Mater. **14**, 871 (2015).

[10] A. R. Mellnik et al., *Spin-Transfer Torque Generated by a Topological Insulator*, Nature **511**, 449 (2014).

[11] A. Manchon, J. Železný, I. M. Miron, T. Jungwirth, J. Sinova, A. Thiaville, K. Garello, and P. Gambardella, *Current-Induced Spin-Orbit Torques in Ferromagnetic and Antiferromagnetic Systems*, Rev. Mod. Phys. **91**, 035004 (2019).

[12] P. Gambardella and I. M. Miron, *Current-Induced Spin–Orbit Torques*, Philos. Trans. R. Soc. Math. Phys. Eng. Sci. **369**, 3175 (2011).

[13] E. Saitoh, M. Ueda, H. Miyajima, and G. Tatara, *Conversion of Spin Current into Charge Current at Room Temperature: Inverse Spin-Hall Effect*, Appl. Phys. Lett. **88**, 182509 (2006).

[14] K. Shen, G. Vignale, and R. Raimondi, *Microscopic Theory of the Inverse Edelstein Effect*, Phys. Rev. Lett. **112**, 096601 (2014).

[15] K. Uchida, S. Takahashi, K. Harii, J. Ieda, W. Koshibae, K. Ando, S. Maekawa, and E. Saitoh, *Observation of the Spin Seebeck Effect*, Nature **455**, 778 (2008).

[16] O. Mosendz, J. E. Pearson, F. Y. Fradin, G. E. W. Bauer, S. D. Bader, and A. Hoffmann, *Quantifying Spin Hall Angles from Spin Pumping: Experiments and Theory*, Phys. Rev. Lett. **104**, 046601 (2010).

[17] J.-C. Rojas-Sánchez, N. Reyren, P. Laczkowski, W. Savero, J.-P. Attané, C. Deranlot, M. Jamet, J.-M. George, L. Vila, and H. Jaffrès, *Spin Pumping and Inverse Spin Hall Effect in Platinum: The Essential Role of Spin-Memory Loss at Metallic Interfaces*, Phys. Rev. Lett. **112**, 106602 (2014).

[18] J. C. Rojas-Sánchez et al., *Spin to Charge Conversion at Room Temperature by Spin Pumping into a New Type of Topological Insulator: $\ensuremath{\alpha}$-Sn Films*, Phys. Rev. Lett. **116**, 096602 (2016).

[19] M. B. Jungfleisch, Q. Zhang, W. Zhang, J. E. Pearson, R. D. Schaller, H. Wen, and A.

Hoffmann, *Control of Terahertz Emission by Ultrafast Spin-Charge Current Conversion at Rashba Interfaces*, Phys. Rev. Lett. **120**, 207207 (2018).

[20] T. Seifert et al., *Efficient Metallic Spintronic Emitters of Ultrabroadband Terahertz Radiation*, Nat. Photonics **10**, 483 (2016).

[21] Y. Wu, M. Elyasi, X. Qiu, M. Chen, Y. Liu, L. Ke, and H. Yang, *High-Performance THz Emitters Based on Ferromagnetic/Nonmagnetic Heterostructures*, Adv. Mater. **29**, 1603031 (2017).

[22] C. Zhou et al., *Broadband Terahertz Generation via the Interface Inverse Rashba-Edelstein Effect*, Phys. Rev. Lett. **121**, 086801 (2018).

[23] D. Go, D. Jo, H.-W. Lee, M. Kläui, and Y. Mokrousov, *Orbitronics: Orbital Currents in Solids*, Europhys. Lett. **135**, 37001 (2021).

[24] S. Ding et al., *Harnessing Orbital-to-Spin Conversion of Interfacial Orbital Currents for Efficient Spin-Orbit Torques*, Phys. Rev. Lett. **125**, 177201 (2020).

[25] D. Go and H.-W. Lee, *Orbital Torque: Torque Generation by Orbital Current Injection*, Phys. Rev. Res. **2**, 013177 (2020).

[26] D. Go, D. Jo, C. Kim, and H.-W. Lee, *Intrinsic Spin and Orbital Hall Effects from Orbital Texture*, Phys. Rev. Lett. **121**, 086602 (2018).

[27] D. Go, D. Jo, T. Gao, K. Ando, S. Blügel, H.-W. Lee, and Y. Mokrousov, *Orbital Rashba Effect in a Surface-Oxidized Cu Film*, Phys. Rev. B **103**, L121113 (2021).

[28] H. Kontani, T. Tanaka, D. S. Hirashima, K. Yamada, and J. Inoue, *Giant Orbital Hall Effect in Transition Metals: Origin of Large Spin and Anomalous Hall Effects*, Phys. Rev. Lett. **102**, 016601 (2009).

[29] D. Lee et al., *Orbital Torque in Magnetic Bilayers*, Nat. Commun. **12**, 1 (2021).

[30] S. Lee et al., *Efficient Conversion of Orbital Hall Current to Spin Current for Spin-Orbit Torque Switching*, Commun. Phys. **4**, 1 (2021).

[31] T. Tanaka, H. Kontani, M. Naito, T. Naito, D. S. Hirashima, K. Yamada, and J. Inoue, *Intrinsic Spin Hall Effect and Orbital Hall Effect in 4 d and 5 d Transition Metals*, Phys. Rev. B **77**, 165117 (2008).

[32] Q. Zhang, Z. Luo, H. Li, Y. Yang, X. Zhang, and Y. Wu, *Terahertz Emission from Anomalous Hall Effect in a Single-Layer Ferromagnet*, Phys. Rev. Appl. **12**, 054027 (2019).

[33] Y. Liu, H. Cheng, Y. Xu, P. Vallobra, S. Eimer, X. Zhang, X. Wu, T. Nie, and W. ZHAO, *Separation of Emission Mechanisms in Spintronic Terahertz Emitters*, Phys. Rev. B **104**, 064419 (2021).

[34] E. Beaurepaire, G. M. Turner, S. M. Harrel, M. C. Beard, J. Bigot, and C. A. Schmuttenmaer, *Coherent Terahertz Emission from Ferromagnetic Films Excited by Femtosecond Laser Pulses*, Appl. Phys. Lett. **84**, 3465 (2004).

[35] T. S. Seifert, D. Go, H. Hayashi, R. Rouzegar, F. Freimuth, K. Ando, Y. Mokrousov, and T. Kampfrath, *Time-Domain Observation of Ballistic Orbital-Angular-Momentum Currents with Giant Relaxation Length in Tungsten*, arXiv:2301.00747.

[36] G. Sala and P. Gambardella, *Giant Orbital Hall Effect and Orbital-to-Spin Conversion in 3d, 5d, and 4f Metallic Heterostructures*, Phys. Rev. Res. **4**, 033037 (2022).

[37] P. Wang et al., *Inverse Orbital Hall Effect and Orbitronic Terahertz Emission Observed in the Materials with Weak Spin-Orbit Coupling*, Npj Quantum Mater. **8**, 1 (2023).

[38] D. Go, F. Freimuth, J.-P. Hanke, F. Xue, O. Gomonay, K.-J. Lee, S. Blügel, P. M. Haney,

H.-W. Lee, and Y. Mokrousov, *Theory of Current-Induced Angular Momentum Transfer Dynamics in Spin-Orbit Coupled Systems*, Phys. Rev. Res. **2**, 033401 (2020).

[39] C Stamm, C Murer, MS Wrnle, Y Acremann, and P Gambardella, *X-Ray Detection of Ultrashort Spin Current Pulses in Synthetic Antiferromagnets*, J. Appl. Phys. (2020).

[40] E. Santos, J. E. Abrão, D. Go, L. K. de Assis, J. B. S. Mendes, and A. Azevedo, *Inverse Orbital Torque via Spin-Orbital Entangled States*.

[41] S. Krishnia, Y. Sassi, F. Ajejas, N. Reyren, S. Collin, A. Fert, J.-M. George, V. Cros, and H. Jaffres, *Large Interfacial Rashba Interaction and Giant Spin-Orbit Torques in Atomically Thin Metallic Heterostructures*, arXiv:2205.08486.

## Acknowledgements

The authors gratefully acknowledge the National Key Research and Development Program of China (No. 2022YFB4400200), National Natural Science Foundation of China (No. 92164206, 62105011, 11904016, 52261145694 and 52121001), Beihang Hefei Innovation Research Institute Project (BHKX-19-01, BHKX-19-02). We acknowledge the ANR program ORION through Grant No. ANR-20-CE30-0022-02. All the authors sincerely thanks Hefei Truth Equipment Co., Ltd for the help on film deposition. This work was supported by the Tencent Foundation through the XPLORER PRIZE.

## Author information

### Authors and Affiliations

**MIIT Key Laboratory of Spintronics, School of Integrated Circuit Science and Engineering, Beihang University, Beijing, China**

Yong Xu, Yongshan Liu, Renyou Xu, Yuhao Jiang, Houyi Cheng, Weisheng Zhao

**Hefei Innovation Research Institute, Beihang University, Hefei, China**

Yong Xu, Fan Zhang, Yongshan Liu, Renyou Xu, Houyi Cheng, Weisheng Zhao

**Unité Mixte de Physique, CNRS, Thales, Université Paris-Saclay, Palaiseau 91767, France**
Albert Fert, Henri-Yves Jaffres

### Contributions

W.Z. initialized, conceived and supervised the project. W.Z. Y.X. F.Z. and R.X. conceived the experiments. Y.L. and H.C. deposited the heterostructures. F.Z. carried out the THz measurements. A.F. Y.X. and H.J. analyzed the experimental data. Y.X. F.Z. H.J. and A.F. wrote the manuscript together. All authors discussed the results and commented on the manuscript. These authors contributed equally: Yong Xu, Fan Zhang, Albert Fert, Henri-Yves Jaffres.
Corresponding author: albert.fert@cnrs.fr, weisheng.zhao@buaa.edu.cn

## DATA AVAILABILITY:

Source data are provided with this paper. All other data that support the finding of this work are available from the corresponding authors upon reasonable request.